\documentclass[aip, amsmath,amssymb, reprint]{revtex4-1}

\usepackage{color}
\usepackage[colorlinks = true,
            linkcolor = blue,
            urlcolor  = blue,
            citecolor = blue,
            anchorcolor = blue]{hyperref}

\usepackage{graphicx}
\usepackage{dcolumn}
\usepackage{bm}

\usepackage[]{siunitx}
\usepackage[utf8]{inputenc}
\usepackage[T1]{fontenc}
\usepackage{mathptmx}
\usepackage{etoolbox}

\makeatletter
\def\@email#1#2{%
 \endgroup
 \patchcmd{\titleblock@produce}
  {\frontmatter@RRAPformat}
  {\frontmatter@RRAPformat{\produce@RRAP{*#1\href{mailto:#2}{#2}}}\frontmatter@RRAPformat}
  {}{}
}%
\makeatother
\begin{document}

\title[On-chip calibrated radio-frequency measurement at cryogenic temperatures for determination of SrTiO$_3$-based capacitor properties]{On-chip calibrated radio-frequency measurement at cryogenic temperatures for determination of SrTiO$_3$-based capacitor properties}

\author{Akitomi Shirachi}
\affiliation{Research Institute of Electrical Communication, Tohoku University, 2-1-1 Katahira, Aoba-ku, Sendai 980-8577, Japan}
\affiliation{Department of Electronic Engineering, Graduate School of Engineering, Tohoku University, Aoba 6-6-05, Aramaki, Aoba-Ku, Sendai 980-8579, Japan}

\author{Motoya Shinozaki}
\affiliation{WPI Advanced Institute for Materials Research, Tohoku University, 2-1-1 Katahira, Aoba-ku, Sendai 980-8577, Japan}

\author{Yasuhide Tomioka}
\affiliation{National Institute of Advanced Industrial Science and Technology (AIST), 1-1-1 Higashi, Tsukuba 305-8565, Japan}

\author{Hisashi Inoue}
\affiliation{National Institute of Advanced Industrial Science and Technology (AIST), 1-1-1 Higashi, Tsukuba 305-8565, Japan}

\author{Kenta Itoh}
\affiliation{Faculty of Science and Engineering, Waseda University, 3-4-1 Okubo, Shinjuku-ku, Tokyo 169-8555, Japan}
\affiliation{Research Center for Materials Nanoarchitechtonics (MANA), National Institute for Material Science (NIMS),
1-2-1 Sengen, Tsukuba 305-0047, Japan}

\author{Yusuke Kozuka}
\affiliation{Faculty of Science and Engineering, Waseda University, 3-4-1 Okubo, Shinjuku-ku, Tokyo 169-8555, Japan}
\affiliation{Research Center for Materials Nanoarchitechtonics (MANA), National Institute for Material Science (NIMS),
1-2-1 Sengen, Tsukuba 305-0047, Japan}

\author{Takanobu Watanabe}
\affiliation{Faculty of Science and Engineering, Waseda University, 3-4-1 Okubo, Shinjuku-ku, Tokyo 169-8555, Japan}

\author{Shoichi Sato}
\affiliation{WPI Advanced Institute for Materials Research, Tohoku University, 2-1-1 Katahira, Aoba-ku, Sendai 980-8577, Japan}

\author{Takeshi Kumasaka}
\affiliation{WPI Advanced Institute for Materials Research, Tohoku University, 2-1-1 Katahira, Aoba-ku, Sendai 980-8577, Japan}

\author{Tomohiro Otsuka}
\email{tomohiro.otsuka@tohoku.ac.jp}
\affiliation{Research Institute of Electrical Communication, Tohoku University, 2-1-1 Katahira, Aoba-ku, Sendai 980-8577, Japan}
\affiliation{Department of Electronic Engineering, Graduate School of Engineering, Tohoku University, Aoba 6-6-05, Aramaki, Aoba-Ku, Sendai 980-8579, Japan}
\affiliation{WPI Advanced Institute for Materials Research, Tohoku University, 2-1-1 Katahira, Aoba-ku, Sendai 980-8577, Japan}
\affiliation{Center for Science and Innovation in Spintronics, Tohoku University, 2-1-1 Katahira, Aoba-ku, Sendai 980-8577, Japan}
\affiliation{RIKEN Center for Emergent Matter Science, 2-1 Hirosawa, Wako, Saitama 351-0198, Japan}

\date{\today}

\begin{abstract}
Quantum computing has emerged as a promising technology for next-generation information processing, utilizing semiconductor quantum dots as one of the candidates for quantum bits.
Radio-frequency (rf) reflectometry plays an important role in the readout of quantum dots but requires a precise rf measurement technique at cryogenic temperatures. 
While cryogenic calibration techniques, essential for rf reflectometry, have been developed, on-chip calibration near the device remains an important challenge.
In this study, we develop an on-chip calibrated rf measurement system operating at \qty{4}{\kelvin} for characterizing SrTiO$_3$-based varactors, which are promising components for tunable impedance matching circuits. 
Our system enables accurate measurements by eliminating errors associated with long rf circuit lines. 
We investigate the effects of annealing conditions, crystal orientation, and Ca doping of SrTiO$_3$ crystals on the varactor properties in the frequency range for rf reflectometry. 
Our results provide insights for optimizing these components for cryogenic rf applications in quantum information processing systems.
\end{abstract}

\maketitle

\begin{figure*}
\begin{center}
  \includegraphics{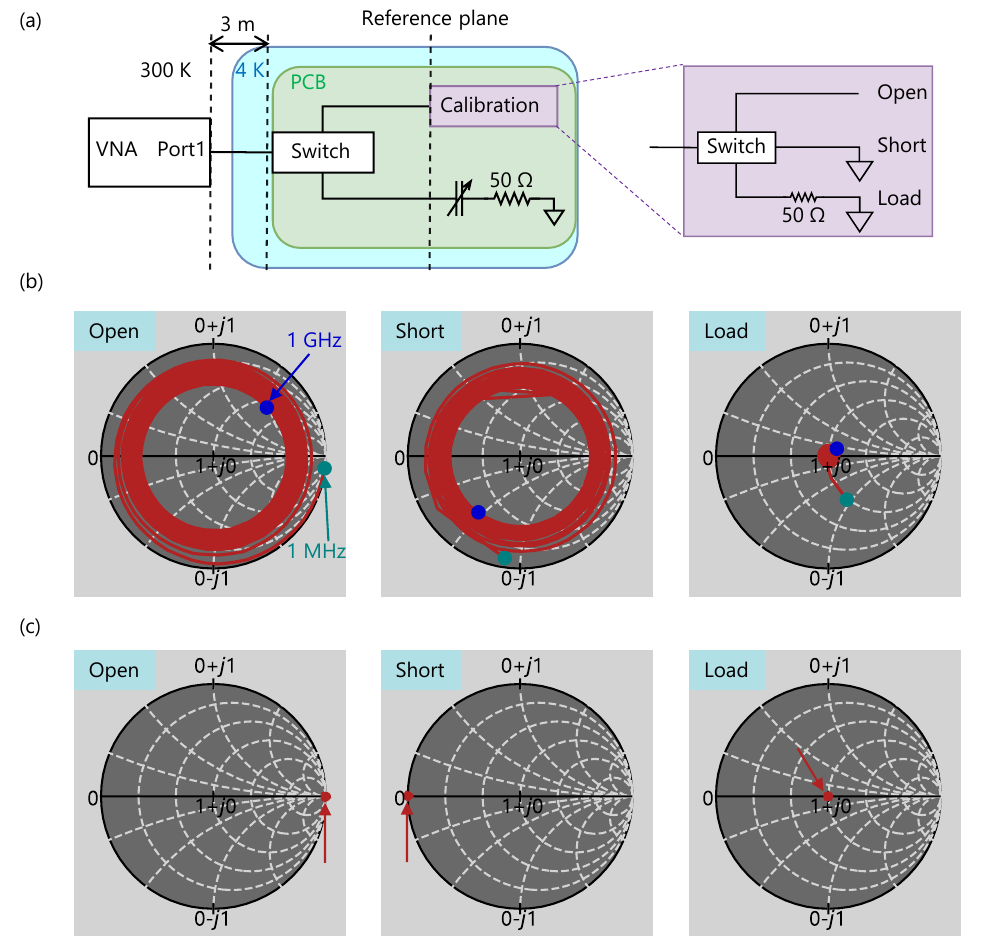}
  \caption{(a) Experimental setup of the calibrated measurement system utilizing HEMT switches on a PCB to divide circuit lines between calibration and measurement ports. 
  (b) Smith charts for open, short, and load conditions at the PCB calibration port while calibrated at the vicinity of the VNA. 
  (c) Smith charts after calibration at the PCB port, showing ideal responses for all conditions.}
  \label{fig1}
\end{center}
\end{figure*}

Quantum computers have attracted attention as next-generation information processing systems. 
Semiconductor quantum dots are considered one of the promising candidates for their essential building blocks, quantum bits (qubits)~\cite{morton2011hybrid, vandersypen2017interfacing}. 
Radio frequency (rf) technologies play an important role in controlling~\cite{koppens2005control} and reading~\cite{qin2006radio, reilly2007fast, barthel2009rapid} the states of semiconductor qubits.
Since semiconductor quantum dots operate at cryogenic temperatures, various components of the rf circuits must be placed on the cryogenic stages of the refrigerator. 
These components are required to exhibit their expected characteristics in the driving frequency range under cryogenic conditions~\cite{al2013cryogenic, Howe2022digital, pauka2021cryogenic, grytsenko2024characterization}.
While it is crucial to evaluate their characteristics at cryogenic temperature, many commercial components are supplied with their datasheets calibrated at room temperature or over \qty{100}{\kelvin}.
Moreover, calibration techniques inside the refrigerator are not yet well established, which means many commercial components cannot promise their functionality at cryogenic temperatures.
The performance of rf components at cryogenic temperatures can be verified through functional circuits including quantum devices. 
This leads to a challenge for designing measurement systems when components show unexpected behavior under cryogenic conditions.

To address this issue, several calibration techniques have been developed using a coaxial switch that operates over a wide temperature range, from cryogenic to room temperatures~\cite{yeh2013situ, perez2023cryogenic, arakawa2023cali, arakawa2025determination}.
These techniques perform calibration at the end of the circuit where the components will be placed for accurate characterisation.
For example, a calibrated rf measurement of up to \qty{26.5}{\giga\hertz} has been reported using this system~\cite{arakawa2023cali}, enabling the evaluation of a circulator's characteristics at cryogenic temperatures.
These advances highlight the growing demand for calibrated measurement systems at cryogenic temperatures, as they are important for the reliable characterization of components in their actual operating environment.

For semiconductor quantum dots, rf reflectometry is a well-established measurement technique commonly referred to as broadband measurement~\cite{qin2006radio, reilly2007fast, barthel2009rapid, vigneau2023probing}. 
This method employs an LC resonator including quantum dots, where capacitance plays an important role in determining the resonator characteristics.
This resonator transforms a device resistance into \qty{50}{\ohm} to satisfy an impedance matching condition with the rf circuit line.
While rf reflectometry has been applied to quantum dots in some material systems such as GaAs, Si~\cite{Liu2021Radio}, and ZnO~\cite{noro2025charge} owing to various efforts of engineering, this technique is still difficult to apply to high-resistance systems such as silicon metal oxide semiconductor-based quantum dots~\cite{bohuslavskyi2024scalable, Tsoukalas2024Pro} and two-dimensional materials like graphene~\cite{banszerus2021dispersive, johmen2023radio, Reuckriegel_electric2024, shinozaki2025rfsoc} due to their high contact resistance~\cite{Li2023Approach}.
Varactor capacitors, externally controlled capacitors, are expected to optimize the resonator characteristics even when we employ high-resistance systems, enabling high-sensitivity reading~\cite{ares2016sensitive, apostolidis2024quantum}.

Strontium titanate (SrTiO$_3$) has emerged as a promising candidate for cryogenic varactor materials.
It is a quantum paraelectric material that maintains a high dielectric constant at cryogenic temperatures without transitioning to a ferroelectric state~\cite{neville1972permittivity}, making it ideal for tunable capacitive applications.
While previous studies have demonstrated its potential as a varactor~\cite{apostolidis2024quantum}, including its robustness to magnetic fields~\cite{Eggli2023Cryo}, its dielectric properties at cryogenic temperature have only been experimentally evaluated at frequencies much lower than those typically used in rf reflectometry~\cite{neville1972permittivity} or specific frequency of a resonator~\cite{Eggli2023Cryo}.
To evaluate rf-dependent characteristics of the varactor, we should calibrate on a printed circuit board (PCB).
This calibration is performed at a reference plane close to device components on the PCB, such as a capacitor, and results in more accurate characterization.
In this study, we develop an on-chip cryogenic calibration circuit to evaluate the properties of SrTiO$_3$-based varactors in the radio frequency regime.

\begin{figure}
\begin{center}
  \includegraphics{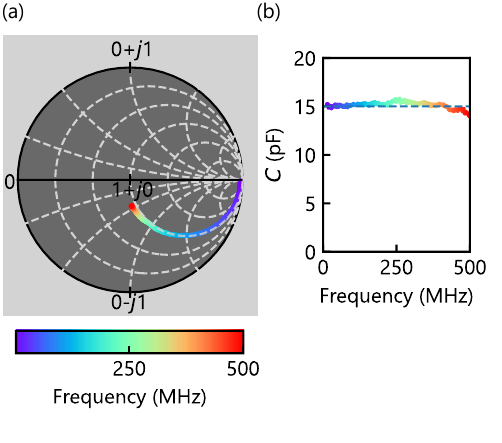}
  \caption{(a) Smith chart for a \qty{15}{\pico\farad} capacitor with frequency sweep from 1 to \qty{500}{\mega\hertz}.
  (b) Frequency dependence of the calculated capacitance from the Smith chart.}
  \label{fig2}
\end{center}
\end{figure}

Figure~\ref{fig1}(a) shows an experimental setup of our calibrated measurement system.
We use high electron mobility transistor (HEMT) switches on the PCB to divide circuit lines to calibration and measurement ports.
The calibration circuit is constructed by open, short, and load ports, which are also divided by HEMT switches.
A circuit length from the vector network analyzer (VNA) to the PCB is approximately \qty{3}{\meter}.
This coaxial cable length affects the measurement accuracy due to frequency-dependent phase shifts.
These circuit line effects would introduce errors in the characterization of components at cryogenic temperatures, especially at higher frequencies where wavelengths become comparable to the circuit scales.
Furthermore, a reference plane is set close to the measurement port to enable a more accurate evaluation.
Figure~\ref{fig1}(b) shows Smith charts under open, short, and load conditions at the PCB calibration port while they are calibrated in the vicinity of the VNA.
All conditions show unexpected trajectories with sweeping the frequency ranging from \qty{1}{\mega\hertz} to \qty{1}{\giga\hertz}.
After calibration using the PCB calibration port, we observe ideal results with all conditions as illustrated in Fig.~\ref{fig1}(c).
This calibration is also useful at cryogenic temperatures.

As a reference, we measure a capacitor with a known capacitance of \qty{15}{\pico\farad} at \qty{4}{\kelvin}.
Figure~\ref{fig2}(a) shows the Smith chart with sweeping the frequency ranging from \qty{1}{} to \qty{500}{\mega\hertz}.
The trajectory of normalized impedance appears along the \qty{50}{\ohm} circle in the lower region, indicating a capacitive component.
From the Smith chart, we obtain a frequency dependence of the capacitance value as shown in Fig.~\ref{fig2}(b).
Owing to our calibration system, the capacitance is evaluated to be \qty{15}{\pico\farad} and maintains its value across the entire frequency range.
This frequency range covers that typically used in rf reflectometry for resistive readout~\cite{vigneau2023probing}.

\begin{figure}
\begin{center}
  \includegraphics{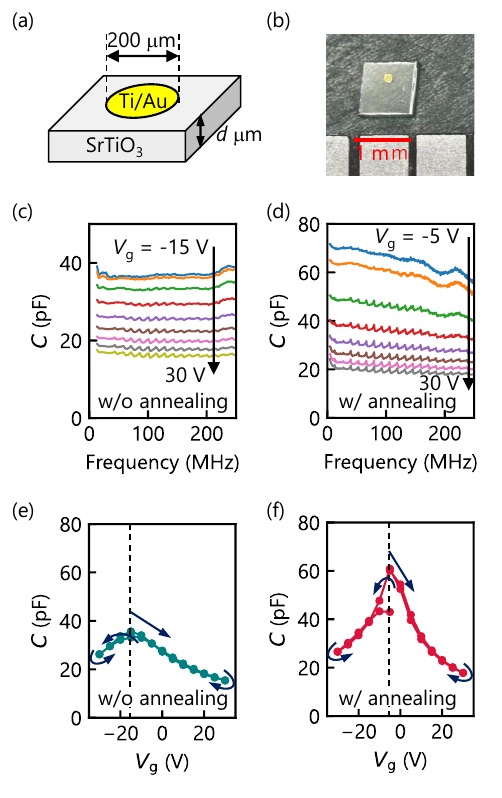}
  \caption{(a) Schematic illustration of the varactor structure with a Ti/Au circular electrode deposited on a SrTiO$_3$ crystal. 
  (b) Top view optical image of a typical device with dimensions of approximately \qty{1}{\milli\meter}. 
  Frequency dependence of capacitance for (c) un-annealed and (d) annealed (110) SrTiO$_3$ devices at \qty{4}{\kelvin} under various gate voltages.
  Gate voltage dependence of capacitance at \qty{200}{\mega\hertz} for (e) un-annealed and (f) annealed devices.
  }                                                 
  \label{fig3}
\end{center}
\end{figure}

Figure~\ref{fig3}(a) illustrates the structure of the SrTiO$_3$-based varactor. 
Here, SrTiO$_{3}$ single crystals are grown by the Verneuil process (Shinkosha Co.).
A Ti/Au layer is deposited on a SrTiO$_3$ crystal and processed into a circular electrode with a diameter of \qty{200}{\micro\meter} by photolithography.
A top view of a typical device is shown in Fig.~\ref{fig3}(b).
We shape devices into squares with dimensions of approximately \qty{1}{\milli\meter}.

At first, we prepare two devices, where one is annealed for \qty{30}{\hour} at 1250 $^\circ$C in the air environment with thickness $d$ of \qty{330}{\micro\meter} and the other is not annealed with $d$ of \qty{260}{\micro\meter}.
Both crystals have the (110) orientation and are not Ca-doped.
Figures~\ref{fig3}(c) and (d) show the frequency dependences of the capacitance from our calibrated measurement at \qty{4}{\kelvin}.
For the device without annealing, the capacitance remains almost constant with frequency up to \qty{250}{\mega\hertz} under all gate voltage $V_\mathrm{g}$ conditions.
The device with annealing shows larger capacitance than that of the un-annealed one, even though it is thicker.
Annealing is considered to fill the residual oxygen vacancy and thus exhibit a higher dielectric constant~\cite{hoshina2018effect}.
Note that the capacitance of the annealed device appears to decrease with increasing frequency.
One of the possible reasons of this decay might result from the difference between the actual value of series resistance to the capacitor and the reference value of \qty{50}{\ohm}.
The total impedance of the measurement port can be described as $R+1/j\omega C$, where $R$ is the series resistance, $j$ the imaginary unit, and $C$ the capacitance of the device.
When this actual resistance deviates from the \qty{50}{\ohm} reference used in the calculation, the calculated capacitance value is underestimated at higher frequencies where the resistive component is dominant over the total impedance.
Therefore, such underestimation is more pronounced in the higher capacitance region shown in Fig.~\ref{fig3}(d), consistent with our model. 
Even after considering this scenario, intrinsic frequency dependence still appears to remain.
Further investigations are necessary to understand this decay in detail.

We extract the capacitance values at \qty{200}{\mega\hertz}, typically used in rf reflectometry, and show the $V_\mathrm{g}$ dependence of both devices in Figs.~\ref{fig3}(e) and (f).
Both cases show clear $V_\mathrm{g}$ modulation without noticeable hysteresis during roundtrip sweeping, which is attributed to the SrTiO$_3$ remaining in the paraelectric state at cryogenic temperature.
The peak of the capacitance shows an offset from zero bias due to the asymmetric top and back electrodes as we employ Ag paste on the back of the device to be grounded.
Therefore, the difference in work functions between the Ti/Au top electrode and Ag back electrode creates an effective internal electric field~\cite{patai1951contact}, causing the offset.

\begin{figure}
\begin{center}
  \includegraphics{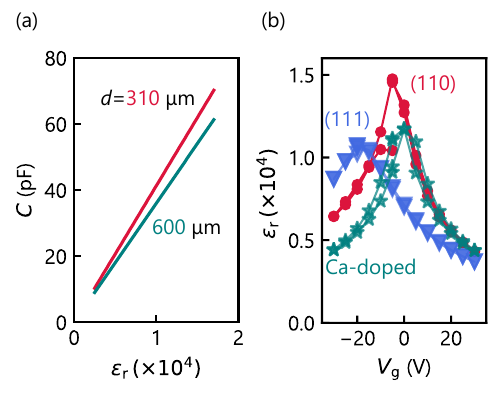}
  \caption{(a) The relationships between the capacitance and relative permittivity for different substrate thicknesses, estimated by COMSOL simulation. 
  (b) Gate voltage dependence of the relative permittivity for devices with different crystal orientations (110), (111), and Ca doping.
  }
  \label{fig4}
\end{center}
\end{figure}

In order to investigate the potential of SrTiO$_3$ varactors in the rf frequency region, we measure the crystal orientation and Ca doping dependence.
For a fair comparison accounting for device geometry, especially the substrate thickness, we perform simulations using COMSOL Multiphysics\textregistered~to convert the measured capacitance $C$ to relative permittivity $\varepsilon_\mathrm{r}$, as shown in Fig.~\ref{fig4}(a).
Using these calculated relationships, we compare $\varepsilon_\mathrm{r}$ between devices with SrTiO$_3$ (110), SrTiO$_3$ (111), and Ca-doped SrTiO$_3$ (110).
Here, the thickness $d$ of the non-doped devices is \qty{310}{\micro\meter} while that of the doped one is \qty{600}{\micro\meter}, and the doped Ca concentration is 0.0015.
Note that the Ca-doped device is grown by the floating zone method and annealed for \qty{30}{\hour} at 1350 $^\circ$C in an Ar/H$_2$ environment, while non-doped devices at 1250 $^\circ$C in air.
Figure~\ref{fig4}(b) summarizes the $V_\mathrm{g}$ dependence of the $\varepsilon_\mathrm{r}$ of each device.
The device with the non-doped (110) crystal exhibits higher $\varepsilon_\mathrm{r}$ values compared to the (111) crystal, consistent with previous reports on the anisotropic dielectric properties of SrTiO$_3$~\cite{Muller1979STO, chang2005plane}.
Regarding the effect of Ca doping, we observe that the Ca-doped device shows reduced $\varepsilon_\mathrm{r}$ values compared to its non-doped counterpart.
This reduction might result from oxygen vacancies and interfacial dielectric characteristics between the substrate and Ag paste, whereas an enhancement in $\varepsilon_\mathrm{r}$ would be expected at low frequencies~\cite{Bednorz1984SrTiO}.
We also observe slight hysteresis during $V_\mathrm{g}$ sweeping, suggesting a ferroelectric transition at this doping concentration and temperature.

In this study, we have developed the on-chip calibrated radio-frequency measurement system operating at cryogenic temperatures for the determination of SrTiO$_3$-based varactor properties. 
Our calibration technique, utilizing HEMT switches on a PCB, enables accurate impedance measurements directly at the device location at \qty{4}{\kelvin}, eliminating errors associated with long transmission lines.
Our setup provides reliable capacitance values across the frequency range using rf reflectometry. 
By using this system, we have investigated dependencies on annealing conditions, crystal orientation, and Ca doping effects of SrTiO$_3$ varactors.
These findings contribute to the understanding of SrTiO$_3$-based varactors in the rf frequency range at cryogenic temperatures, providing essential insights for their application in quantum device measurements. 
The calibration technique developed in this study can be further applied to characterize various cryogenic microwave components such as superconducting inductors, supporting the advancement of quantum information processing systems.

\begin{acknowledgments}
The authors thank A. Kurita, RIEC Fundamental Technology Center, and the Laboratory for Nanoelectronics and Spintronics for technical support.
Part of this work was supported by
MEXT Leading Initiative for Excellent Young Researchers, 
Grants-in-Aid for Scientific Research (21K18592, 22H04958, 23K26482, 23H04490), 
FRiD Tohoku University, and by TIA “KAKEHASHI” program.
AIMR and MANA are supported by World Premier International Research Center Initiative (WPI), MEXT, Japan.
\end{acknowledgments}

\section*{AUTHOR DECLARATIONS}
\subsection*{Conflict of Interest}
The authors have no conflicts to disclose.
\subsection*{Author Contributions}
\textbf{Akitomi Shirachi:} Data Curation (lead); Investigation (lead); Methodology (equal); Visualization (equal); Writing/Review \& Editing (equal). 
\textbf{Motoya Shinozaki:} Conceptualization (equal); Data Curation (equal); Investigation (equal); Methodology (equal);  Visualization (lead);; Writing/Original Draft (lead); Writing/Review \& Editing (equal);
\textbf{Yasuhide Tomioka:} Investigation (equal); Resources (lead); Writing/Review \& Editing (equal);
\textbf{Hisashi Inoue:} Investigation (equal); Resources (equal); Writing/Review \& Editing (equal);
\textbf{Kenta Itoh:} Investigation (equal); Software (lead); Writing/Review \& Editing (equal);
\textbf{Yusuke Kozuka:} Conceptualization (equal); Investigation (equal); Resources (equal); Software (equal); Writing/Review \& Editing (equal);
\textbf{Takanobu Watanabe:} Investigation (equal); Software (equal); Writing/Review \& Editing (equal);
\textbf{Shoichi Sato:} Methodology (equal); Resources (equal); Writing/Review \& Editing (equal);
\textbf{Takeshi Kumasaka:} Resources (equal); Writing/Review \& Editing (equal);
\textbf{Tomohiro Otsuka:} Conceptualization (lead); Methodology (lead); Funding Acquisition (lead); Supervision (lead); Writing/Review \& Editing (lead).

\section*{Data Availability Statement}
The data that support the findings of this study are available from the corresponding authors upon reasonable request.

\bibliography{reference.bib}

\end{document}